\documentstyle[12pt]{article}

\def\V{\rule{0pt}{2.5ex}}           
\def\D{\rule[-1ex]{0pt}{1ex}}       
\def\bigfrac#1#2{\frac{\V#1}{\D#2}}
\newcommand{\bea}{\begin{eqnarray}}
\newcommand{\eea}{\end{eqnarray}}
\newcommand{\apart}{\left( \begin{array}{l}}
\newcommand{\cpart}{\end{array} \right) }
\newcommand{\aparg}{\left\{ \begin{array}{l}}
\newcommand{\cparg}{\end{array} \right\} }
\newcommand{\aparq}{\left[ \begin{array}{l}}
\newcommand{\cparq}{\end{array} \right] }
\newcommand{\fraz}{\displaystyle\bigfrac}
\newcommand{\tl}{\theta(\lambda)}
\newcommand{\th}{\theta}
\newcommand{\laj}{\lambda_j}
\newcommand{\lak}{\lambda_k}
\newcommand{\nj}{n_j}

\newcommand{\DN}{\Delta N}
\newcommand{\dpa}{\partial}
\newcommand{\QB}{\overline{Q}}

\newcommand{\VB}{\overline{V}}
\newcommand{\nb}{\overline{n}}
\newcommand{\beq}{\begin{equation}}
\newcommand{\eeq}{\end{equation}}
\newcommand{\ovl}{\overline}
\newcommand{\wtl}{\widetilde}

\newcommand{\dl}{\delta}
 
\newcommand{\lam}{\lambda}
\newcommand{\p}{^{\prime}}

\begin{document}
\begin{flushright}
DFTT 27/97\\
May 1997
\end{flushright}
\begin{center}
{\Large \bf Extended conformal symmetry of the one--dimensional Bose gas}
\vskip 0.8cm
Milena~MAULE,~~Stefano~SCIUTO\footnote[1]{e-mail: sciuto@to.infn.it}\\
\vskip 0.3cm
{\em Dipartimento di Fisica Teorica, Universit\`a di Torino,\\
and I.N.F.N. Sezione di Torino\\
Via P. Giuria 1, 10125 Torino, Italy}
\end{center}
\vskip 1cm
\begin{abstract}
We show that the low--lying excitations of the one--dimensional Bose gas are 
described, at all orders in a $1/N$ expansion and at the first order in 
the inverse of the coupling constant, by an effective hamiltonian 
written in terms of an extended conformal algebra, namely the Cartan 
subalgebra of the $W_{1+\infty}\times \ovl{W}_{1+\infty}$ algebra.
This enables us to construct the first interaction term which corrects the
hamiltonian of free fermions equivalent to a hard--core boson system.
\end{abstract}
\vskip 1cm
It is well known \cite{hald, frahm, kawa} that the
low--lying excitations of all one--dimensional gapless models solvable by 
Bethe Ansatz can be described by an effective hamiltonian
\beq
\label{4}
{\cal H}\propto \fraz{1}{N} (L_0+\ovl{L}_0),
\eeq
where $N$ is the number of particles and $L_0$ and $\ovl{L}_0$ are the zero
modes of the right and left Virasoro (conformal) algebra respectively
\cite{bela}.
Recently \cite{frau} it has been shown that also the subleading 
terms in a $1/N$
expansion of the Heisenberg and Calogero--Sutherland models (and of a large
class of fermionic models) have an algebraic description in terms of the
$W_{1+\infty}\times \ovl{W}_{1+\infty}$ algebra, which is a linear and infinite
dimensional extension of the Virasoro algebra. 

The prototype of the models solvable by Bethe Ansatz is 
the one--dimensional Bose gas \cite{bethe} with hamiltonian:
\beq
\label{1}
H_N=\sum\limits_{j=1}^N \apart -\fraz{\dpa^2}{\dpa x_j^2} \cpart +
2c\sum\limits_{N\geq j>k\geq 1}
\delta(x_j-x_k);~~~c>0,~x_j \in [0,L].
\eeq
Since for large $c$ the Bose gas
is equivalent to a gas of free fermions, it is plausible that the 
$W_{1+\infty}\times\ovl{W}_{1+\infty}$ algebra also describes the subleading
terms of the effective hamiltonian of this system; 
in this letter we shall show that this is so at
the first order in $1/c$. In the literature \cite{berko1, berko2}
a $1/N$ expansion of the effective hamiltonian has already been considered, but
there the irrelevant terms were 
written in terms of primary fields of the conformal algebra; here we shall show
that they can be interpreted as generators of the Cartan subalgebra of $W_{1+
\infty}\times \ovl{W}_{1+\infty}$.

The energy spectrum of the Bose gas is simply given by:
\beq
\label{sp}
E=\sum\limits_{j=1}^{N} \laj^2 ,
\eeq
where the pseudo--particle momenta $\laj$ satisfy the Bethe equations:
\bea
\label{A}
L \laj + \sum\limits_{k=1}^N \th(
\laj -\lak)=2\pi I_j;
\eea
$\tl$ is a monotonically increasing function of $\lam$:
\bea
\label{B}
\tl= i \ln \apart \fraz{ic +\lam}{ic- \lam}\cpart
\eea
and the $I_j$'s are integers (if $N$ is odd) or half--integers (if $N$ is even)
which uniquely parameterize the solutions of the Bethe equations \cite{yang}.

Specifically, the set of numbers $I^{0}_{j}$ given by:
\bea
\label{C}
I^{0}_{j}= - \fraz{N-1}{2}+j-1,~~~~j=1,\ldots,N
\eea
represents the ground state of the system.
The numbers $I^0_1=-(N-1)/2$ and $I^0_N=(N-1)/2$ are called left and right Fermi
points respectively.

Let us now consider the low--lying excitations above the ground state that have
zero energy in the thermodynamic limit ($N \rightarrow \infty, L \rightarrow 
\infty, \rho_0=N/L=$constant). They can be obtained in three different
ways:
\begin{itemize}
\item by adding $\DN$ particles to the system (with $|\DN|\ll N$);
\item by moving $d$ particles from the left to the right Fermi point
(with $|d|\ll N$);
\item by creating particle--hole pairs at the points $(I_j + n_{N-j+1},~ I_j)$
and $(I_j-\nb_j,~ I_j)$ on
the right and left respectively (with $n_j, \nb_j \neq 0$ only if $j\ll N$ and
$0\leq n_j, \nb_j \ll N$).
\end{itemize}
A generic low--lying excitation, obtained by combining these three processes, 
is labelled by the following quantum numbers:
\beq
\label{D}
I_{j}=\wtl{I}^{0}_{j} + d -\overline{n}_j +n_{N\p -j+1},~~~~~~j=1,2,\ldots ,N\p
\eeq
where
\beq
\label{E}
\wtl{I}^{0}_{j}=-\fraz{N\p -1}{2} +j-1,~~~~~~~~j=1,2,\ldots ,N\p
\eeq
with $N\p=N+\DN$ and the integers $n_j, \nb_j$ ordered according to:
\begin{displaymath}
n_1\geq n_2\geq \ldots \geq 0,~~~~~~\nb_1\geq \nb_2\geq\ldots \geq 0.
\end{displaymath}

We observe that when the coupling constant $c$ tends to infinity, the particles
of the system behave like free fermions, and the Bethe equations can be solved
immediately \cite{gira, lieb}. 
We thus choose $c$ large but finite and solve the Bethe equations at the 
first order in $1/c$, in order to calculate the pseudo--particle
momenta and therefore the energy of the system.

In this way we obtain the excitation energy - related
to the three physical processes described above - at the first order in $1/c$
and at all orders in the powers of $1/N$:
\bea
\label{5}
&\dl E&=(2\pi\rho_0)^2 \Bigg\{
\fraz{1}{4}\Bigg(1-\fraz{8}{3}g\Bigg)\DN+
\fraz{1}{N}\Bigg[\fraz{1}{4}(1-4g)\DN^2+\nonumber\\
&+&d^2+(1-2g)(n+\overline{n})\Bigg]+
\fraz{1}{N^2}\Bigg[\fraz{1}{12}(1-8g)\DN^3+\nonumber\\
&-&\fraz{1}{12}
(1-4g)\DN+d^2\DN+
(1-2g)\Bigg(\sum\limits_{j}(n_j^2+\overline{n}_j^2)+\\
&+&(\DN+1)(n+\overline{n})-2\sum\limits_j j(\nj+\overline{n}_j)\Bigg)
-2g\DN(n+\overline{n})+\nonumber\\
&+&2d(n-\overline{n})\Bigg]
-\fraz{2g}{N^3}
\Bigg[\fraz{1}{12}(\DN^4-\DN^2)+\DN \Bigg(\sum\limits_{j}(n_j^2+\overline{n}_j^2)+
\nonumber\\
&+&(\DN+1)(n+\overline{n})-2\sum\limits_j j(\nj+\overline{n}_j)\Bigg)
-(n-\overline{n})^2\Bigg]\Bigg\} + O(g^2)\nonumber ,
\eea
where:\\
$g=\rho_0 2/c $, $n=\sum n_j$ and 
$\overline{n}=\sum \overline{n}_j$.

To take into account the $O(1/N^2)$ correction terms, the effective
hamiltonian (\ref{4}) must also contain irrelevant terms 
which we shall show to be generators of the
$W_{1+\infty}\times \ovl{W}_{1+\infty}$ algebra. 
The $W_{1+\infty}$ algebra is generated by an infinite set of currents 
$V^i_n$, characterized
by a mode index $n \in \mbox{\boldmath $Z$}$ and an integer conformal spin
$h=i+1\geq 1$. These currents satisfy the algebra \cite{bakas}:
\beq
\label{6}
[V_n^i,V_m^j]=(jn-im)V_{n+m}^{i+j-1}+q(i,j,n,m)
V_{n+m}^{i+j-3}+\ldots+\dl^{ij}\dl_{n+m,0} C d(i,n),
\eeq
where $C$ is the central charge (that we choose equal to one), the structure 
constants $q,\ldots ,d$ are polynomial in
their arguments, and the dots denote a finite number of terms involving the
operators $V_{n+m}^{i+j-2k}$. We explicitly write some of the equations
(\ref{6}) for $C=1$:
\bea
\label{7}
&[V_n^0,V_m^0]&= n \dl_{n+m,0}\nonumber\\
&[V_{n}^{1},V_{m}^{0}]&=-mV_{n+m}^{0}\nonumber\\
&[V_{n}^{1},V_{m}^{1}]&=(n-m)V_{n+m}^{1}+\fraz{1}{12}n(n^2 -1)\dl_{n+m,
0}\nonumber\\
&[V_{n}^{2},V_{m}^{0}]&=-2mV_{n+m}^{1}\\
&[V_{n}^{2},V_{m}^{1}]&=(n-2m)V_{n+m}^{2}-\fraz{1}{6}(m^3-m)
V_{n+m}^{0}\nonumber\\
&[V_{n}^{2},V_{m}^{2}]&=(2n-2m)V_{n+m}^{3}+\fraz{n-m}{15}(2n^2+2m^2-nm-8)
V_{n+m}^{1}+\nonumber \\
&~~~+& \fraz{n(n^2-1)(n^2-4)}{180}\dl_{n+m,0}\nonumber ,
\eea
which show that the operators $V^0_n$ satisfy the
Abelian Kac--Moody algebra $\widehat{U(1)}$, and the operators $V^1_n$ satisfy
the Virasoro algebra \cite{bela}. 
The operators $\ovl{V}^i_n$ close the same algebra and commute with the
$V^i_n$'s.

$W_{1+\infty}$ algebra representations are built from the highest weight states
$|Q>$, defined by:
\begin{displaymath}
V_0^i |Q>=f^i(Q) |Q>
\end{displaymath}
\beq
\label{8}
V_n^i |Q>=0,~~~~~\forall~ n>0,~\forall~ i\geq 0,
\eeq
where $f^i (Q)$ are polynomials in $Q$ determined by the algebra. The
complete highest weight representation is obtained by constructing all the
descendant states of $|Q>$ \cite{kac}:
\beq
\label{9}
|Q,\{k_i\}>=V^0_{-k_{1}}V^0_{-k_{2}}\ldots V^0_{-k_{s}}|Q>,~~~k_1\geq k_2\geq
\ldots\geq k_s > 0.
\eeq

~From the algebra (\ref{7}), the
eigenvalues of the operators $V^i_0$ on the descendant states can be calculated
obtaining:
\beq
\label{22}
V^0_0|Q,\{n_i\}>=Q|Q,\{n_i\}>,
\eeq
\beq
\label{24}
V^1_0|Q,\{n_i\}>=\Bigg[\fraz{Q^2}{2}+\sum\limits_{i=0}^{s}n_i\Bigg]|Q,\{n_i\}>,
\eeq
\beq
\label{25}
V^2_0|Q,\{n_i\}>=
\aparq \fraz{Q^3}{3} +2Q\sum\limits_{i=0}^s n_i +\sum\limits_{i=0}^s \Big(n_i^
2-(2i-1)n_i\Big)\cparq |Q,\{n_i\}>
\eeq
and so on. Analogous equations apply for the $\VB^i_0$ operator eigenvalues 
on $|\QB,\{ \nb_i\}>$.

We can thus write an effective hamiltonian in terms of the Cartan subalgebra of
$W_{1+\infty}\times \overline{W}_{1+\infty}$ algebra, which generalizes
(\ref{4}) to all orders of a $1/N$ expansion:
\bea
\label{21}
&{\cal H}=&(2\pi\rho_0)^2\Bigg\{\fraz{1}{4}\Bigg(1-\fraz{5}{3}g\Bigg)(V^0_0+\VB^0_0)+
\fraz{1}{N}\Bigg[(1-2g)(V^1_0+\VB^1_0)\Bigg]+\nonumber\\
&+&\fraz{1}{N^2}\Bigg[(1-2g)(V^2_0+\VB^2_0)-\fraz{1-3g}{12}(V^0_0+\VB^0_0)-
2g(V^0_0 \VB^1_0+\VB^0_0 V^1_0)\Bigg]+\\
&-&\fraz{2g}{N^3}\Bigg[(V^2_0+\VB^2_0)(V^0_0+\VB^0_0)-(V^1_0-\VB^1_0)^2-
\fraz{1}{12}(V^0_0+\VB^0_0)^2\Bigg]\Bigg\}+O(g^2)\nonumber .
\eea
In fact the eigenvalues of ${\cal H}$ on the descendant states of the type 
$|Q,\QB, \{n_i\}, \{\nb_i\}>$ reproduce the excitation energy (\ref{5}), 
calculated by means of the Bethe Ansatz, provided that $Q$ and $\QB$ denote 
the quantities:
\beq
\label{27}
\begin{array}{ll}
Q=&\fraz{\DN}{2\sqrt{Z}}+\sqrt{Z} d\nonumber\\
\QB=&\fraz{\DN}{2\sqrt{Z}}-\sqrt{Z} d,
\end{array}
\eeq
where $\sqrt{Z}=1+g$ can be interpreted as the compactification radius of a 
free boson field in terms of which all the $C=1~~$ $W_{1+\infty}\times\ovl{W}_
{1+\infty}$ algebra representations can be built \cite{bela}.

In this way we have proved that the complete effective hamiltonian of the
system displays a $W_{1+\infty}$ structure: its Hilbert space is described by a
set of unitary, irreducible, highest weight representations of the
$W_{1+\infty}\times\ovl{W}_{1+\infty}$ algebra, which are known and completely 
classified \cite{kac}.

As already noted, when the coupling constant $c$ tends to infinity our model is
equivalent to a fermionic one. Since there is also a fermionic realization of
the $C=1$ $W_{1+\infty}\times \ovl{W}_{1+\infty}$ algebra, we shall be able to
construct the interaction term
that corrects, at the first order in $1/c$, the hamiltonian of
free fermions equivalent to a hard--core boson system. 
However the fermionic representation is equivalent 
to a bosonic one only when the
compactification radius is equal to one. The compactification radius can be
changed by defining a new set of operators $W^0_l$, $\ovl{W}^0_l$ which still 
satisfy the Kac--Moody algebra in (\ref{7}) and $[W^0_l, \ovl{W}^0_m]=0,~\forall
l,m$:
\beq
\label{Z}
\begin{array}{lll}
W^0_0=V^0_0 \cosh \beta + \ovl{V}^0_0\sinh \beta,&~~~~W^0_l=V^0_l&~~l\neq 
0\nonumber\\
\ovl{W}^0_0=\ovl{V}^0_0\cosh \beta + V^0_0\sinh \beta,&~~~~\ovl{W}^0_l =
\ovl{V}^0_l&~~l\neq 0
\end{array}
\eeq
with $\tanh \beta=\fraz{Z-1}{Z+1}=g+O(g^2)$.
The eigenvalues of $W^0_0$ and $\ovl{W}^0_0$ on a descendant state are now
given by (\ref{27}) with $\sqrt{Z}$ replaced by one. A new $W_{1+\infty}\times
\ovl{W}_{1+\infty}$ algebra corresponding to unitary compactification radius
can then be built by defining $W^i_l$, $\ovl{W}^i_l$ ($i \geq 1,l \in
\mbox{\boldmath $Z$}$) in terms of $W^0_m$, $\ovl{W}^0_m$ by means of a
generalized Sugawara construction \cite{kac}. Then the hamiltonian (\ref{21})
can be expressed in terms of this new $W_{1+\infty}\times\ovl{W}_{1+\infty}$
algebra by a combined use of the generalized Sugawara construction and of
(\ref{Z}); we obtain:
\bea
\label{29}
{\cal H}&=&(2\pi\rho_0)^2 \Bigg\{ (1-2g)\fraz{N}{12} +\fraz{1}{4} \Bigg(
1-\fraz{8}{3}g\Bigg)(W^0_0+\ovl{W}^0_0)+\nonumber\\
&+&\fraz{1}{N} \Bigg[ (1-2g)\Bigg( W^1_0+\ovl{W}^1_0 -\fraz{1}{12} \Bigg) -2g
W^0_0\ovl{W}^0_0 \Bigg]+\\
&+&\fraz{1}{N^2}\Bigg[ (1-2g)(W^2_0+\ovl{W}^2_0)
-4g(W^0_0\ovl{W}^1_0+\ovl{W}^0_0 W^1_0) -\fraz{1}{12} (1-4g)(W^0_0+\ovl{W}^0_0)
\Bigg]+\nonumber\\
&-&\fraz{2g}{N^3} \Bigg[ (W^2_0+\ovl{W}^2_0)(W^0_0+\ovl{W}^0_0)
-(W^1_0-\ovl{W}^1_0)-\fraz{1}{12}(W^0_0+\ovl{W}^0_0)^2\Bigg]\Bigg\}
+ O(g^2)\nonumber .
\eea
The fermionic representation of the $W_{1+\infty}\times\ovl{W}_{1+\infty}$
algebra enables us to write 
the operators $W^i_n,\ovl{W}^i_n$ as fermionic bilinear operators;
following the procedures described in \cite{frau}, it is possible to see that 
the effective hamiltonian (\ref{29}) corresponds to a second quantized 
hamiltonian of the form:
\bea
\label{P}
H&=& \int\limits_0^L dx~ \Psi^{\dag}(x) \apart-\fraz{\dpa^2}{\dpa x^2}\cpart
\Psi(x)+\nonumber\\
&+&\fraz{2}{cL}\int\limits_0^L dx\int\limits_0^L dy \Psi^{\dag}(x)
\Psi^{\dag}(y) \apart \fraz{\dpa}{\dpa x}-\fraz{\dpa}{\dpa y}\cpart^2 \Psi(x)
\Psi(y),
\eea
where $\Psi$ is a fermionic field with
standard anticommutation relations.

On the other hand, it is easy to see that solving the Bethe equations at the
first order in $1/c$ and calculating the energy of the system, we obtain the
following hamiltonian:
\beq
\label{O}
H=\sum_{j=1}^N p_j^2-\fraz{2}{cL}\sum\limits_{j,k}(p_j-p_k)^2.
\eeq
(where $p_j=\fraz{2\pi}{L} n_j$ represent the fermion momenta), which is the
exactly the first quantized version of the hamiltonian (\ref{P}).
\vskip 2.5cm
\noindent
{\large {\bf Acknowledgements}}
\vskip 0.5cm
\noindent
We would like to thank G. R. Zemba for pointing out reference \cite{berko2}
and M. Frau, A. Lerda and R. Caracciolo for useful discussions.
\skip 1.5cm

\end{document}